\documentclass[usenatbib]{mn2e}

\usepackage{dcolumn}
\usepackage{graphics,epsfig}
\usepackage{txfonts}
\usepackage{times}
\usepackage{amssymb}

\def\hi{H{\sc i}}

\newcommand{\kms}{km~s$^{-1}$}
\newcommand{\cm}{cm$^{-2}$}

\newcommand{\noi}{\noindent}
\newcommand{\lb}{\left(}

\newcommand{\rb}{\right)}

\newcommand{\ts}{T_s}
\newcommand{\tx}{T_x}
\newcommand{\beq}{\begin{equation}}
\newcommand{\eeq}{\end{equation}}

\title[Outflowing atomic and molecular gas at $z \sim 0.67$]{Outflowing atomic and molecular gas at $z \sim 0.67$ towards 1504+377}
\author[Kanekar \& Chengalur]{Nissim~Kanekar$^1$\thanks{E-mail: nkanekar@aoc.nrao.edu (NK); chengalu@ncra.tifr.res.in (JNC)} 
Jayaram~N.~Chengalur$^{2}$\\
$^1${}National Radio Astronomy Observatory, 1003 Lopezville Rd, Socorro, NM 87801, USA; \\
$^2${}National Centre for Radio Astrophysics, Ganeshkhind, Pune--411007, India}

\begin{document}
\date{Received mmddyy/ accepted mmddyy}
\maketitle
\label{firstpage}

\begin{abstract}

We report the detection of OH~1667~MHz and wide \hi~21cm absorption at $z \sim 0.67$ towards the 
red quasar 1504+377, with the Green Bank Telescope and the Giant Metrewave Radio Telescope. 
The \hi~21cm absorption extends over a velocity range of $\sim 600$~km/s blueward of 
the quasar redshift ($z=0.674$), with the new OH~1667~MHz absorption component at 
$\sim -430$~\kms, nearly coincident with earlier detections of mm-wave absorption at 
$z \sim 0.6715$. The atomic and molecular absorption appear to arise from a fast gas 
outflow from the quasar, with a mass outflow rate ${\dot M} \sim 12 M_\odot$~yr$^{-1}$ 
and a molecular hydrogen fraction $f_{\rm H_2} \equiv (N_{\rm H_2}/N_{\rm HI}) \sim 0.2$. 
The radio structure of 1504+377 is consistent with the outflow arising due to a jet-cloud 
interaction, followed by rapid cooling of the cloud material. The observed ratio of 
HCO$^+$ to OH column densities is $\sim 20$~times higher than typical values in 
Galactic and high-$z$ absorbers. This could arise due to small-scale structure in 
the outflowing gas on sub-parsec scales, which would also explain the observed
variability in the \hi~21cm line.

\end{abstract}


\section{Introduction}
\label{intro}

The quasar 1504+377 is a rare case of a radio-loud active galactic nucleus (AGN) hosted by a disk
galaxy (e.g. \citealt{perlman96,carilli97}). The flat-spectrum radio emission arises from 
a compact core and a one-sided jet to the southwest, with the jet axis aligned (within 
$\sim 15^\circ$) with the major axis of the host galaxy \citep{polatidis95,fomalont00}. 
The AGN is heavily reddened ($r$-K~=~5.1) and was not detected in a deep R-band image, 
suggesting a high level of dust obscuration \citep{stickel96}. Consistent with 
this, strong redshifted mm-wave molecular absorption has been detected towards the 
radio source \citep{wiklind96}, with two absorption complexes at $z \sim 0.6734$ (system~A) 
and $z \sim 0.6715$ (system~B), close to the redshift of the host galaxy 
($z = 0.674 \pm 0.001$; \citealt{stickel94}).

Besides the mm-wave transitions, \hi~21cm, OH~1665~MHz and OH~1667~MHz absorption 
have all been detected from system~A, with strong, wide profiles extending over a 
velocity range of $\gtrsim 100$~\kms\ \citep{wiklind96,carilli97,kanekar02}. In contrast, 
the mm-wave absorption in system~B is quite narrow [full-width-at-half-maximum (FWHM)$ \sim 15$~\kms]
and neither \hi~21cm nor OH absorption have been detected at this redshift 
\citep{carilli97,carilli98}.  
This is the only $z > 0.1$ mm-wave absorber that has not hitherto been detected 
in OH or \hi~21cm absorption 
\citep{wiklind94,wiklind95,wiklind96,wiklind96b,chengalur99,kanekar02,kanekar03c}
and is thus an excellent candidate for a deep search in these transitions. Beside studying 
physical conditions in the interstellar medium (ISM) of the QSO host, the detection 
of these lines would, in principle, also allow one to test for changes in the fundamental 
constants from $z \sim 0.67$ to the present epoch \citep{darling03,chengalur03,kanekar04b}. 
Unfortunately, the OH~1665~MHz line from $z \sim 0.6715$ lies at the same frequency as the 
known 1667~MHz absorption from $z \sim 0.6734$ \citep{kanekar02}, implying that it 
(and the latter 1667~MHz line) cannot be used to probe fundamental constant evolution. 
We report here a search for the other three redshifted OH 
ground-state lines (at rest frequencies of 1667.3590, 1612.2310 and 1720.5299~MHz) and the 
\hi~21cm line towards 1504+377 with the Giant Metrewave Radio Telescope (GMRT) 
and the Green Bank Telescope (GBT), resulting in the detection of OH~1667~MHz and \hi~21cm 
absorption at $z \sim 0.6715$. 

\section{Observations and data analysis}
\label{sec:obs}

\begin{figure*}
\centering
\epsfig{file=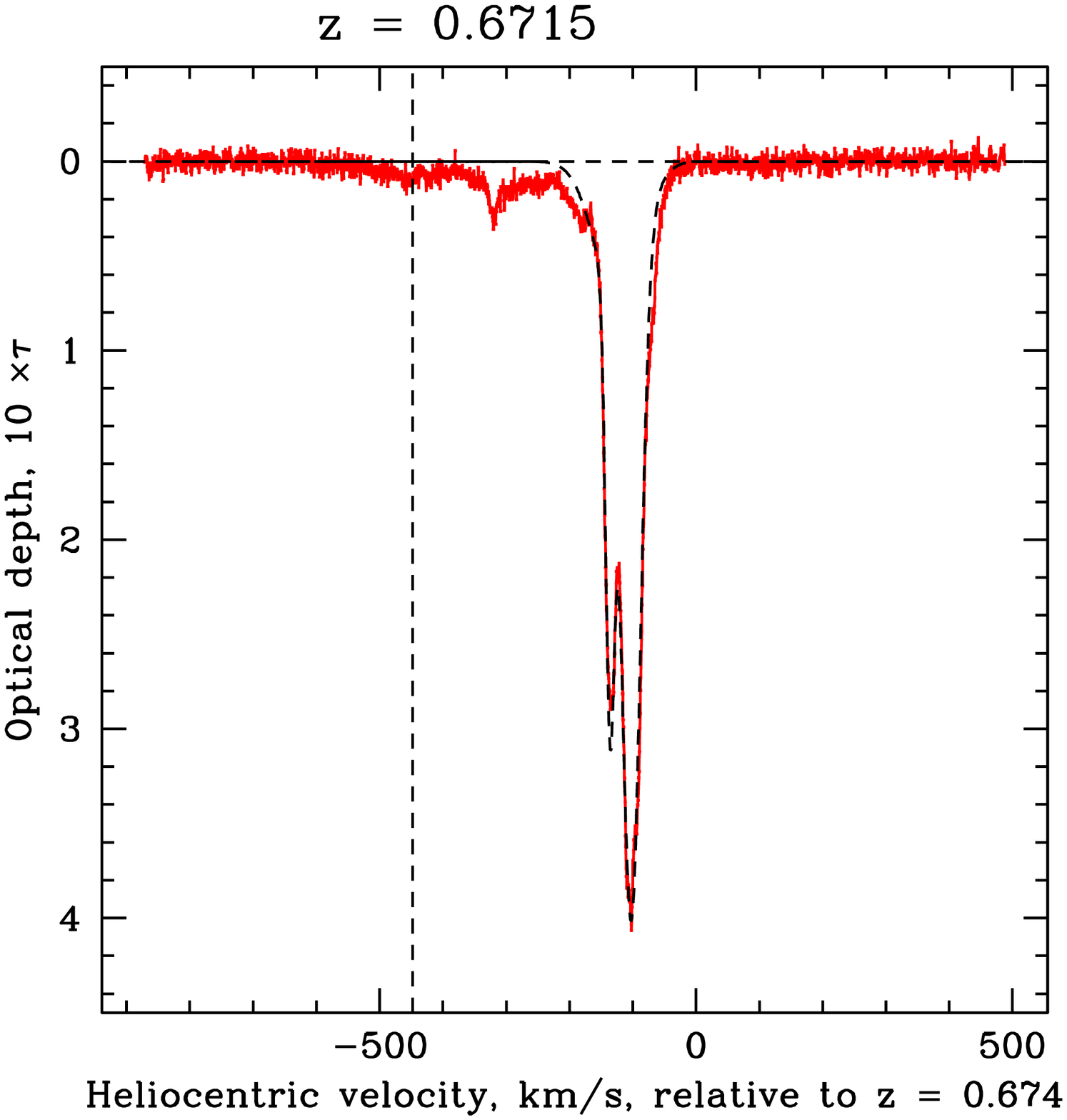,height=3.3truein,width=3.3truein}
\epsfig{file=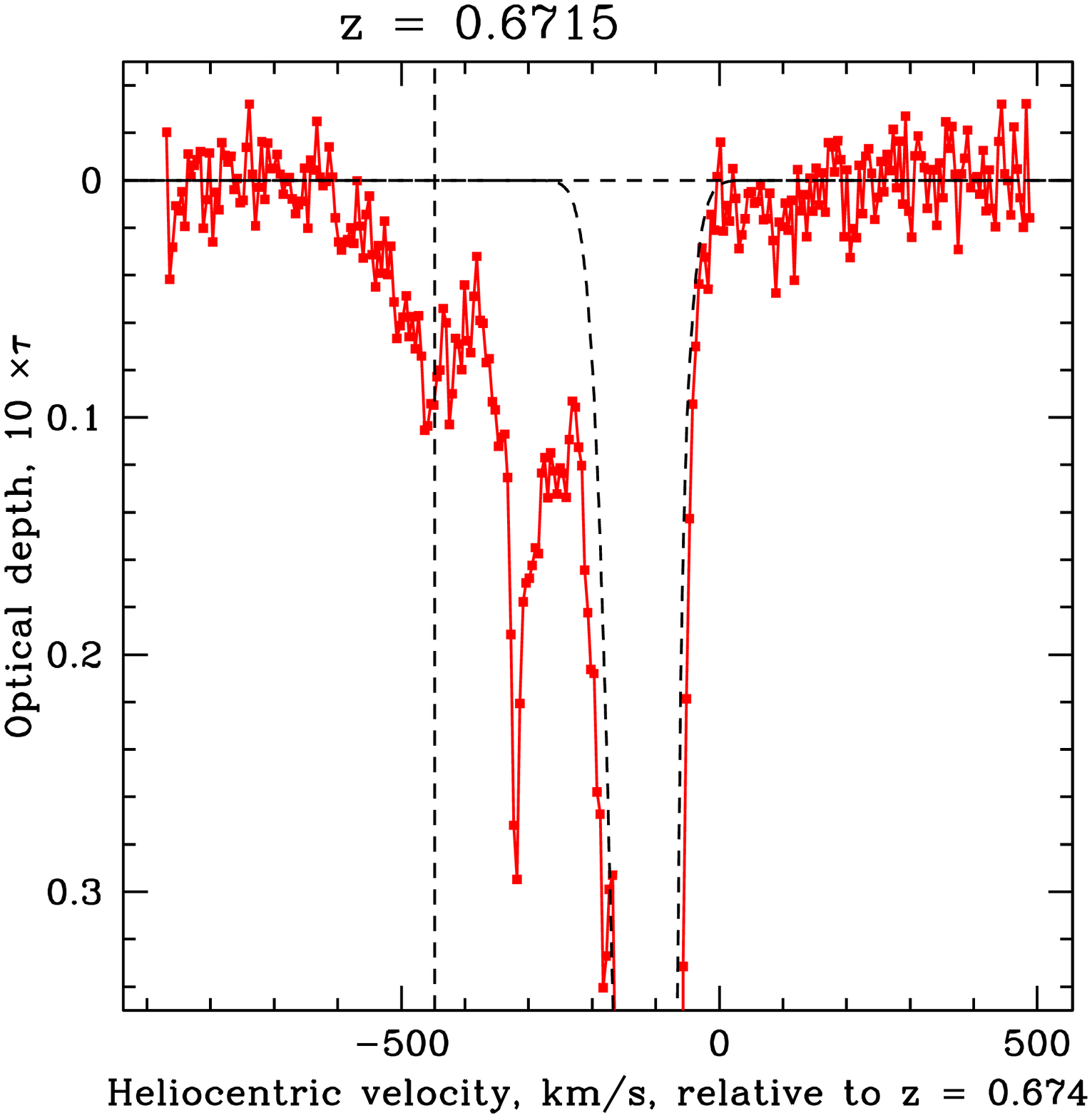,height=3.3truein,width=3.3truein}
\caption{Final GBT \hi~21cm absorption spectrum towards 1504+377 (left panel; resolution $\sim 
0.54$~\kms), with optical depth plotted against heliocentric velocity (relative to $z = 0.674$, 
the quasar redshift). The right panel shows the spectrum smoothed to a resolution of 
$\sim 4.8$~\kms\ and zoomed-in.  The dashed vertical line indicates $z = 0.6715$ while 
the dashed curves show the 3-Gaussian fit of \citet{carilli98} to their WSRT spectrum.} 
\label{fig:21cm}
\vskip -0.1in
\end{figure*}

A search for the 1667~MHz and 1720~MHz OH lines from $z \sim 0.6715$ was 
initially carried out with the GMRT on 26 and 27~March~2006, using the 256-channel 
mode of the correlator. Bandwidths of 1~and 4~MHz, centred at 997.37 and 1028.77~MHz,
were used for the 1667 and 1720~MHz observations, respectively (also allowing a 
search for the 1720~MHz line from system~A), yielding  velocity resolutions 
of $\sim 2.3$~\kms\ and $\sim 9.1$~\kms\ after Hanning smoothing.
3C286 was used to calibrate the flux density scale
and the bandpass shape; no secondary calibrator was observed as 1504+377 is a phase 
calibrator for the GMRT. The on-source times in the 1667 and 1720~MHz transitions 
were $\sim 4.6$~hours and $\sim 1.3$~hours, respectively.

The GMRT data were analysed in ``classic'' AIPS, using standard procedures. After initial 
editing to remove corrupted data, continuum images were made of the field at the 
two frequencies; both images yielded a flux density of $\sim 1.04 \pm 0.01$~Jy for 1504+377. 
The radio continuum at each frequency was then subtracted out using the task UVLIN and 
the residual visibilities shifted to the heliocentric frame and imaged in all channels. 
The final spectra were then extracted by a cut through the spectral cubes at 
the location of 1504+377.

The GMRT observations resulted in the detection of a weak absorption feature at the expected 
frequency of the redshifted 1667~MHz line. To confirm this and to obtain a better \hi~21cm 
spectrum, we retrieved archival GBT datasets covering the redshifted \hi~21cm line (from December~2003) 
and all four ground-state OH lines (from September~2004). The \hi~21cm line was later
re-observed with the GBT in November~2006, to confirm the wide, weak absorption seen in 
the archival data.  

The GBT observations were carried out in total-power, position-switched mode. 
The OH~runs used four 12.5~MHz Auto-Correlation Spectrometer (ACS) bands, with 
8192~channels, centred on the redshifted OH~18cm frequencies. This allowed simultaneous 
coverage of all four OH~18cm lines from both redshifts, with velocity resolutions of 
$\sim 0.9$~\kms\ after Hanning smoothing. The \hi~21cm observations of 2003 and 2006 
used a single 12.5~MHz ACS band with 16384 and 32768 channels, 
respectively, giving resolutions of $\sim 0.54$~\kms\ (in 2003) and $\sim 0.27$~\kms\ 
(in 2006). The on-source times were 2.5~hours for the OH lines and 0.3~and 1.5~hours for 
the \hi~21cm line in 2003 and 2006, respectively.  

All GBT data were analysed in {\sc DISH}, the {\sc AIPS++} single-dish package, using 
standard procedures. After initial data editing and calibration, the continuum flux density 
was measured using RFI- and line-free channels. A second-order baseline was then 
fit to each (typically 10-second) record and subtracted out during the process of calibration; 
the residual data were then averaged together to obtain the final spectra for each transition.
In the case of multiple observing epochs (e.g. the \hi~21cm line), the data from different 
runs were averaged together, after smoothing and interpolating to the same spectral 
resolution and frequency scale. The GBT 1720~MHz dataset was affected by strong 
terrestrial interference (RFI) and will hence not be discussed further. 

\section{Results}
\label{sec:results}

\subsection{Spectra}
\label{sec:spectra}

The left panel of Fig.~\ref{fig:21cm} shows the final GBT \hi~21cm spectrum towards 1504+377, 
with 
optical depth (computed assuming a flux density of 1.04~Jy) plotted as a function of 
heliocentric velocity, in \kms, relative to 
$z = 0.674$. This has a root-mean-square (RMS) noise of 0.0032, in optical depth units, 
per $\sim 0.54$~\kms\ channel. The strong 21cm absorption at $\sim -100$~\kms\ (system~A)
was detected by \citet{carilli97}; the dashed curve shows the 3-Gaussian fit of 
\citet{carilli98} to their WSRT spectrum (which we note, in passing, is $\sim 15$~times 
less sensitive than the GBT spectrum of Fig.~\ref{fig:21cm}). The right panel of the
figure shows a zoomed-in version of the spectrum, smoothed to a resolution 
of $\sim 4.8$~\kms\ to clearly display the wide absorption tail. The \hi~21cm absorption 
extends well beyond the absorption detected by \citet{carilli97,carilli98}, with a 
full-width-between-nulls (FWBN) of $\sim 600$~\kms. The new extended absorption can be 
separated into three distinct parts, a narrow component at $\sim -320$~\kms\ (i.e. 
$z = 0.6722$), a broad feature at $\sim -430$~\kms\ [nearly the same redshift as the 
$z \sim 0.6715$ mm-wave absorption of \citet{wiklind96}] and a smooth weak tail, extending out 
to $-600$~\kms. The integrated \hi~21cm optical depth is 
$\int \tau_{\rm 21} \rm{d}V = (27.20 \pm 0.04)$~\kms, 
with around 15\% of the integrated optical depth in the new components detected here. 

\begin{figure}
\centering
\epsfig{file=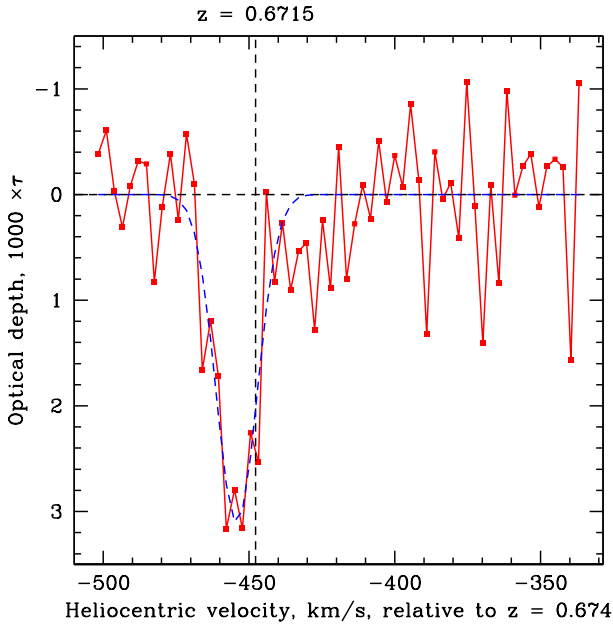,height=3.3truein,width=3.3truein}
\caption{The final redshifted OH~1667~MHz absorption spectrum from system~B towards 
1504+377, with optical depth (in units of $10^3 \times \tau$) plotted against 
heliocentric velocity in \kms, relative to $z = 0.674$.  The redshift $z = 0.6715$ 
is indicated by the dashed vertical line.}
\label{fig:oh1667}
\vskip -0.1in
\end{figure}


Weak narrow absorption was visible close to the expected frequency of the redshifted 
OH~1667~MHz line in both the GMRT and the GBT spectra (at $\sim 5\sigma$ significance 
in each spectrum, after averaging all channels). It is very unlikely that the absorption 
is due to local RFI, given that the spectra were taken at independent telescopes, separated 
by a period of 2.5 years and with very different doppler shifts. Fig.~\ref{fig:oh1667}
shows our final OH~1667~MHz spectrum, obtained by averaging the GMRT and GBT spectra with 
weights determined by the RMS noise values [$\sim 8 \times 10^{-4}$~(GMRT) and $\sim 
1.3 \times 10^{-3}$~(GBT), per 2.75~\kms\ channel, in optical depth units].
The optical depth RMS noise on this spectrum is 
$\sim 6.4 \times 10^{-4}$ per $2.75$~\kms\ channel. The spectrum is well-fit by a 
single Gaussian model (the dashed curve in the figure), with FWHM$ =(16.5 \pm 2.2)$~\kms\ 
and a peak optical depth of $(3.2 \pm 0.4) \times 10^{-3}$, at $z = 0.6714637 (51)$. 
The integrated 1667~MHz optical depth is $\int \tau_{\rm 1667} 
\rm{d}V \sim (0.067 \pm 0.003)$~\kms.

Finally, absorption was not detected in the redshifted OH~1612 and 1720~MHz lines (not shown 
here) at any velocity. The RMS noise on the GMRT 1720~MHz spectrum is $\sim 0.0012$ per 
$\sim 9.1$~\kms\ channel, while that on the GBT 1612~MHz spectrum is $\sim 0.0024$ 
per $\sim 0.92$~\kms\ channel, in optical depth units. The $3 \sigma$ upper limits on the 
velocity-integrated optical depth in the 1720 and 1612~MHz lines are $\sim 0.06$~\kms\ 
and $\sim 0.04$~\kms, after smoothing the GBT spectrum by 11~channels to a resolution 
of $\sim 10.1$~\kms. These limits assume a Gaussian profile with FWHM$ =16.5$~\kms, 
that of the fit to the OH~1667~MHz line. The ratio of the integrated optical depths 
in the 1665~MHz and 1612~MHz lines of system~A is $R > 6.4$, higher than that expected 
($R \sim 5$) for gas in thermal equilibrium. Note that our sensitivity is insufficient 
to detect the satellite lines from system~B, if the gas is in thermal equilibrium. 

\subsection{HI and OH column densities}
\label{sec:column}

For optically thin gas, the \hi\ and OH column densities can be derived from the 
\hi~21cm and OH~1667~MHz absorption profiles using the expressions 

\begin{equation}
N_{\rm HI} = 1.823 \times 10^{18} {\lb {\frac {\ts}{f_{\rm 21}}} \rb} \: \int \tau_{\rm 21}
\: \mathrm{d} V \:\:\:\: {\rm and}
\label{eqn:21cm}
\end{equation}

\begin{equation}
N_{\rm OH} = 2.24 \times 10^{14} {\lb {\frac {\tx }{f_{\rm OH}}} \rb} \: \int \tau_{\rm 1667} \:
 \mathrm{d} V \; ,
\label{eqn:noh}
\end{equation}
where $\ts$ (in K) is the \hi\ spin temperature, $\tx$ (in K), the OH excitation temperature, 
and $f_{\rm 21}$ and $f_{\rm OH}$ are the \hi\ and OH covering factors at the respective 
redshifted line frequencies. $N_{\rm OH}$ and $N_{\rm HI}$ have units of \cm, while the 
integrals are over velocity, in \kms. \citet{carilli97} used 1.6 and 5~GHz VLBI
observations to estimate $f_{\rm 21} = 0.46$, if only the radio core (of angular size 
$\lesssim 1.4$~mas) is covered, and $f_{\rm 21} = 0.74$, if the inner jet is covered out to
$\sim 10$~mas; this would require the absorbing material to have a spatial extent of 
$\gtrsim 10 \: h_{71}^{-1}$~pc and $\gtrsim 70 \: h_{71}^{-1}$~pc, respectively\footnote{We 
use the standard LCDM cosmology, with H$_0 = 71$~\kms~Mpc$^{-1}$, $\Omega_m = 0.27$ and 
$\Omega_\Lambda = 0.73$.}. Typical sizes of Galactic molecular clouds range from 
$\sim 10 - 50$~pc 
\citep{blitz90}, somewhat smaller than the latter value; we will hence assume that at least 
the radio core is covered in both \hi~21cm and OH lines, i.e. $f_{\rm 21} \ge 0.46$ and 
$f_{\rm OH} \ge 0.46$. 

Next, it is not possible to determine either $\ts$ or $\tx$ using only the \hi~21cm or 
OH~1667~MHz absorption profiles.  Spin temperature estimates range from $\sim 100$~K 
in the Galaxy and local and intermediate redshift spiral disks (e.g. \citealt{braun92}) 
to $\gtrsim 1000$~K in high redshift damped Lyman-$\alpha$ systems \citep{kanekar03}. 
Assuming $\ts = 100$~K gives a lower limit to the \hi\ column density. Further, 
following \citet{kanekar02}, we will assume $\tx \sim 10$~K, a typical temperature 
in dark clouds. We then obtain $N_{\rm HI} \ge (1.08 \pm 0.15) \times (\ts/100) (0.46/f_{21}) 
\times 10^{22}$~\cm\ and $N_{\rm OH} \sim (3.26 \pm 0.15) \times (\tx/10) (0.46/f_{\rm OH}) 
\times 10^{14}$~\cm. It should be emphasized that the above \hi\ column density is for 
the entire profile, i.e. is not restricted to the absorption from system~B, while 
the OH column density is merely for this system. Finally, we use the empirical 
relation $N_{\rm H_2} \sim 10^7 \times N_{\rm OH}$ \citep{liszt99} to estimate the 
molecular hydrogen column density to be $N_{\rm H_2} \sim 3.3 \times (\tx/10) (0.46/f_{\rm OH}) 
\times 10^{21}$~\cm\ for system~B. System~A has 
$N_{\rm H_2} \sim 2.3 \times  (\tx/10) (0.46/f_{\rm OH}) \times 10^{22}$~\cm\ \citep{kanekar02}, 
giving a total ${\rm H_2}$ column density of $N_{\rm H_2} \sim 2.6 \times (\tx/10) 
(0.46/f_{\rm OH}) \times 10^{22}$~\cm\ at $z \sim 0.67$.

\section{Discussion}
\label{sec:discuss}

\begin{figure}
\centering
\epsfig{file=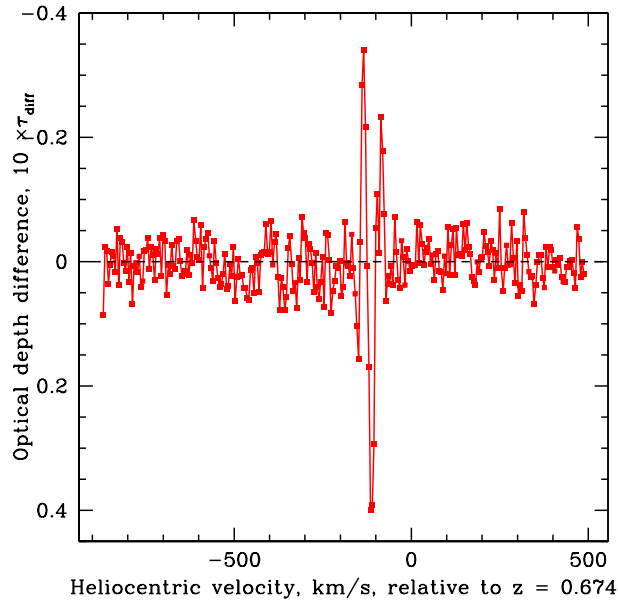,height=3.3truein,width=3.3truein}
\caption{Final difference spectrum between the \hi~21cm optical depth spectra towards 
1504+377 in 2003 and 2006, with optical depth difference (in units of $10 \times \tau_{diff}$) 
plotted against heliocentric velocity in \kms\, relative to $z = 0.674$. The original 
difference spectrum had a resolution of $\sim 0.54$~\kms; this was boxcar-smoothed to, 
and resampled at, a resolution of $\sim 4.8$~\kms\ to produce this spectrum.}
\label{fig:variability}
\vskip -0.1in
\end{figure}

\subsection{Variability in the \hi~21cm profile}
\label{sec:21cm-change}

Fig.~\ref{fig:variability} shows a plot of the difference between the \hi~21cm optical depths
in 2003 and 2006 versus heliocentric velocity, in \kms, relative to $z = 0.674$. The 
strong features in the difference spectrum between $\sim -150$~\kms\ and $\sim -70$~\kms\ 
indicate significant changes ($\sim 10\%$ of the line depth) in the \hi~21cm profile 
between 2003 and 2006. Note that the difference cannot be due to a simple scaling of one or 
both of the spectra, as different spectral components show changes of opposite sign. While
the possibility that the observed change might be due to RFI cannot be ruled out, no
evidence was seen for RFI at these frequencies, in either these or our other 850~MHz 
GBT datasets. The profile ``variability'' is coincident with the strongest spectral 
components, with the rest of the profile showing no evidence for changes within the noise.

Variability in redshifted \hi~21cm profiles has been seen earlier in two damped Lyman-$\alpha$ 
systems, at $z \sim 0.524$ towards 0235+164 \citep{wolfe82} and $z \sim 0.3127$ towards 1127$-$145 
\citep{kanekar01c}. While changes in the latter two profiles have been detected on far
shorter timescales (a few days) than in 1504+377, it is interesting that all three sources 
contain highly compact ($\sim $~mas-scale) components. Possible explanations for the
observed changes towards 1504+377 include refractive scintillation in the Galactic 
interstellar medium  (for which the background source need not be compact; \citealt{macquart05}), 
or transverse motion of a source component on VLBI scales \citep{briggs83c}. Both models 
require small-scale structure in the 21cm optical depth of the absorbing gas.

\subsection{Physical conditions in the absorbing gas}
\label{sec:1504}

The radio core of 1504+377 and the nucleus of the host galaxy are coincident within the errors 
($\sim 1''$) in the R-band image of \citet{stickel94}. At mm-wave frequencies, the core 
dominates the quasar flux density, with very little emission coming from the steep-spectrum 
jet \citep{wiklind96}. The core is also likely to be extremely compact at these frequencies, 
implying that both mm-wave absorbers arise along a single line of sight, which must also pass 
extremely close to the centre of the host galaxy. \citet{wiklind96} noted that it is 
impossible to produce two absorption components at very different velocities in such 
circumstances if the absorbing gas is in pure rotational motion. The large separation 
($\sim 330$~\kms) between the two observed 
absorption velocities is thus suggestive of the presence of strong non-circular orbits; 
these authors argued in favour of a scenario in which the broad absorption from system~A
originates close to the nucleus (in a nuclear ring or a bar), while the narrow absorption 
of system~A arises in a more-distant cloud in the disk of the host galaxy. In this picture, 
the systemic redshift is $z \sim 0.6715$. On the other hand, \citet{carilli97} used the 
fact that the optical emission redshift of the host galaxy ($z = 0.674 \pm 0.001$) is in 
excellent agreement with that of the higher-redshift complex to argue that the latter is the 
systemic redshift. They also pointed out that the outflow velocity of system~A in this model 
($\sim 330$~\kms) is too large for a cloud in the outer disk of the parent galaxy and 
suggested the possibility that it might arise in a high-velocity cloud, due to 
tidal interactions between the host galaxy and a nearby object seen in the R-band image 
of \citet{stickel94}. Our new GBT \hi~21cm spectrum of Fig.~\ref{fig:21cm} clearly 
shows that the two absorption systems are, in fact, part of a continuous absorption 
complex, spanning $\sim 600$~\kms\ and extending from the optical redshift of $z \sim 0.674$ 
out to $z \sim 0.6706$. The 21cm absorption lies entirely blueward of the optical redshift, 
implying that it must arise in gas that is outflowing from the quasar.

The large velocity spread of the \hi\ outflow in 1504+377 is similar to that seen in a number of 
low-redshift AGNs \citep{morganti05}. These authors note that all known fast \hi\ outflows have 
been detected in radio galaxies in early or re-started phases of their radio activity. 
There is also evidence that the most likely mechanism to explain such fast \hi\ outflows 
is interaction between the radio jets and the surrounding interstellar medium (e.g. 
\citealt{morganti05b}), with rapid cooling taking place in the gas after a jet-cloud interaction, 
as expected from numerical simulations (e.g. \citealt{fragile04}). The fact that 1504+377 
shows no extended radio structure (the outer jet extends to only $\sim 55$~mas, i.e. $\sim 387$~pc, 
from the nucleus; \citealt{polatidis95}) suggests that it too is in a early phase of its 
radio activity. Recent 5~GHz VLBI observations \citep{bolton06} have found a new north-eastern 
extension, which was not seen in earlier (deeper) images (e.g. \citealt{fomalont00}), 
demonstrating that the source is currently in an active phase. Finally, the fact that the 
radio structure in 1504+377 is strongly one-sided (e.g. \citealt{fomalont00}) indicates 
that the jet lies close to the line of sight towards the core. The above suggestion that 
jet-cloud interactions are responsible for local gas cooling is consistent with the fact 
that mm-wave absorption (which 
takes place in cold gas and, as noted earlier, must arise towards the core) is seen 
at multiple velocities along the line of sight.

It thus appears that the wide \hi~21cm and molecular absorption towards 1504+377 
arise in outflowing gas from the AGN that is cooling rapidly after an interaction with 
the south-western radio jet. This is the highest redshift at which such a high-velocity 
outflow has been observed (e.g. \citealt{morganti05}) and, perhaps more interesting, the 
first case where molecular gas has been detected in the outflow. The ${\rm H_2}$ fraction 
is $f_{\rm H_2} = \left[N_{\rm H_2}/N_{\rm HI}\right] \le 2 \times (\ts/100) (\tx/10) 
(f_{\rm OH}/f_{\rm 21})$. \citet{morganti05} assume $\ts \sim 1000$~K to estimate 
\hi\ column densities for sources in their sample due to the proximity of the gas 
to the AGN and the likely presence of shocks. Using this value for consistency gives 
a molecular fraction of $f_{\rm H_2} \sim 0.2$ in the outflowing gas.


We estimate the mass outflow rate ${\dot M}$ using the model of \citet{heckman00}, 
in which a constant-velocity, mass-conserving wind flows into a solid angle 
$\Omega$ from a  minimum radius $r_*$, viz.

\begin{equation}
{\dot M} = 30 \left[ \frac{\Omega}{4\Pi}\right] \left[ \frac{r_*}{1 \: {\rm kpc}} \right] 
\left[ \frac{N_{\rm H}}{10^{21} \: {\rm cm^{-2}}} \right] 
\left[ \frac{v}{300 \:{\rm km s^{-1}}} \right] \: M_\odot \: {\rm yr}^{-1},
\end{equation}
\noi where $v$ is the outflow velocity and $N_{\rm H}$, the total hydrogen column density 
of the outflowing gas. We will assume that the minimum radius $r_*$ is $\sim 10$~pc, 
the size of the radio core, and, following \citet{morganti05}, that $\Omega = \Pi$~steradians 
and $v = \:{\rm FWBN}/2 \sim 300$~\kms.  The total hydrogen column density at 
$z \sim 0.67$ is $N_{\rm H} = \left[ N_{\rm HI} + 2 \times N_{\rm H_2} \right] 
\sim 1.6 \times 10^{23}$~\cm, again assuming $\ts \sim 1000$~K. This leads to an 
estimated mass outflow rate of ${\dot M} \sim 12 M_\odot$~yr$^{-1}$, 
comparable to estimates in nearby fast \hi\ outflows \citep{morganti05}.



\citet{wiklind96} noted that HCO$^+$ is highly over-abundant in system~B, enhanced by at 
least an order of magnitude relative to expected abundances in chemical models. The 
ratios of HCO$^+$ to CO and HCO$^+$ to HCN column densities here are $3 - 5$ times 
larger than in system~A. While such large differences in relative abundances between 
HCO$^+$ and species such as CO, HCN, etc, have been 
observed in Galactic clouds \citep{lucas98}, the ratio of OH to HCO$^+$ column densities 
in both the Galaxy and a sample of four redshifted HCO$^+$ and OH absorbers has been 
found to be fairly constant, with $N_{\rm HCO^+}/N_{\rm OH} \sim 0.03$ \citep{liszt96,kanekar02}
over more than two orders of magnitude in HCO$^+$ column density. \citet{liszt04} 
found this ratio to show a large spread (by a factor of $\sim 4$) in the clouds 
towards Cen.A and NGC1052, with the HCO$^+$ and OH lines also showing very different 
kinematics, but argued that this could be explained by differing source structure 
and foreground free-free opacity at the OH and HCO$^+$ frequencies, source variability 
between observing epochs, and excitation effects at high OH column densities ($\gtrsim 
10^{15}$~\cm; \citealt{langevelde95}). Conversely, system~B has 
$N_{\rm HCO^+}/N_{\rm OH} \sim 0.5 \times (\tx/10) (0.46/f_{\rm OH})$, discrepant by more than 
an order of magnitude from the expected value. However, 1504+377 is highly compact at 
both mm-wave and cm-wave frequencies (with a cm-wave core-fraction of $\sim 46$\%; 
\citealt{carilli97}) and the HCO$^+$ and OH lines have very similar FWHMs 
[$\sim 16.5 \pm 2.2$~\kms\ (OH) and $\sim 15.2 \pm 0.9$~\kms\ (HCO$^+$)], making it likely 
that they arise from similar gas. Increasing $\tx$ by an order of magnitude could 
resolve this problem but such high $\tx$ values have never been seen in the Galaxy 
(e.g. \citealt{liszt96}). Similarly, the  ratio of peak optical depths in the 
HCO$^+$ and \hi~21cm lines in system~A is  $R \equiv \tau_{HCO^+}/\tau_{\rm 21} 
\sim 30$, far larger than that seen in Galactic clouds ($0.1 \le R \le 6$; 
\citealt{lucas96,liszt96}). \citet{carilli97} point out that high values of $R$ could 
result from either far warmer \hi\ or low molecular dissociation, but this would not 
explain the discrepancy in the ratio of OH and HCO$^+$ column densities. If the latter 
is not due to real chemical differences between OH and HCO$^+$ (which seems unlikely; 
\citealt{liszt00}), a plausible explanation is extreme small-scale structure in the 
opacity of the absorbing gas on sub-parsec scales, smaller than the size of the radio 
core at cm wavelengths. This could arise due to internal shocks or turbulence in the 
rapidly outflowing gas. As noted earlier, the observed variability in the \hi~21cm absorption 
at $z \sim 0.674$ suggests similar small-scale structure at a different location in the 
outflow, which could also account for the large velocity difference ($\sim 15$~\kms) in peak 
OH and HCO$^+$ absorption in system~A \citep{kanekar02}. 
Monitoring the mm-wave lines for variability would be one way of testing this hypothesis.

Finally, comparisons between the OH, \hi~21cm and HCO$^+$ redshifts from an absorber 
can be used to test the evolution of fundamental constants \citep{darling03,chengalur03}. 
However, the above possibility of small-scale structure in the absorbing gas makes it 
likely that any such comparisons in the absorbing gas towards 1504+377 will be dominated 
by local systematic velocity offsets. We conclude that this absorber is unlikely to be 
useful for the purpose of probing fundamental constant evolution.

%

\section{Acknowledgments}
We thank Francoise Combes for providing us with an ASCII file of the HCO$^+$ spectrum 
towards 1504+377. We also thank the staff of the GMRT who made these observations possible. 
The GMRT is run by the National Centre for Radio Astrophysics of the Tata Institute of 
Fundamental Research. The National Radio Astronomy Observatory is operated by Associated 
Universities, Inc. under cooperative agreement with the National Science Foundation.

\bibliographystyle{mn2e}
\bibliography{ms}

\end{document}